# Responding to the Event Deluge

Roy D. Williams[a], Scott D. Barthelmy[b], Robert B. Denny[c], Matthew J. Graham[a],
and John Swinbank[d]

[a] California Institute of Technology
[b] NASA Goddard Space Flight Center
[c] DC-3 Dreams SP
[d] University of Amsterdam

## ABSTRACT

We present the VOEventNet infrastructure for large-scale rapid follow-up of astronomical events, including selection, annotation, machine intelligence, and coordination of observations. The VOEvent standard is central to this vision, with distributed and replicated services rather than centralized facilities. We also describe some of the event brokers, services, and software that are connected to the network. These technologies will become more important in the coming years, with new event streams from Gaia, LOFAR, LIGO, LSST, and many others.

## 1. Deluge

In the late 1990s, the gamma-ray burst (GRB) community ignited the current excitement over transient astronomical events. Gamma-Ray Bursts (GRBs) were a real enigma until ultra-fast event dissemination allowed optical identification of afterglows, leading to rich data and rich science. The events back then were both valuable and infrequent: every new GRB could make a career for a young astronomer, and they were only detected every few days. However, in the next few years, surveys carried out by telescopes such as Gaia, LOFAR, Pan-STARRS, LSST and SKA will produce a flood of hundreds of events every 24 hours, with the scientific jewels surrounded by dross, and so both scalability and discrimination will be increasingly important. We should also point out that a deluge of event *metadata* is also coming, as more transient surveys come online, and each of those spawns its own streams of follow-up events and annotations.

Follow-up observation will be in short supply in the era of the event deluge. Faint objects can only be observed with the largest telescopes — that are already over-subscribed, and objects with uncertain position require deep wide-field imaging to look for possible counterparts. But selection of "interesting" events is subject to the same quality measures of any selection process, being the false-positive and false-negative rates. False positives are uninteresting things that waste valuable follow-up resources, and false negatives are exciting objects that were not identified as such. It will be important to bring together all possible information, as quickly as possible, and to build new ways of automated information fusion, in order to minimize both false positives and false negatives.

There have been several ways to represent and communicate astronomical transient notices. The Central Bureau of Astronomical Telegrams[1] has been sending transient notices since 1882, and the Astronomer's Telegram website[2] has been running for several years. However these are natural language text, requiring a human in the loop for decision and action. Given the event deluge, and the importance of rapid follow-up, human readable text must give way to communications that can be understood and acted on by machines. The original GCN[3] used formally-defined 160-byte packets, which enabled the blossoming of GRB science as noted above. However, in the last few years the VOEvent [4][5] syntax, from the International Virtual Observatory Alliance[6], has become the common language for rapidly disseminating machine-readable astronomical transients.

Once events are being disseminated, there may be other capabilities that can increase the science harvest from an event stream. Annotation is the process of adding information to a first detection, based on follow-up observation, archival search, or intelligence assessment. Consumers of events may wish to select events based on many kinds of criteria, depending on the event itself, or the associated annotations, or on the ability of their follow-up telescope, or other queries. Those building machine-intelligence codes or mining events will want a repository (database) of events and their annotations, to build training sets, test their software, or decide on selection criteria.

In this paper, we introduce VOEventNet, a network architecture for broadcasting VOEvents, and we discuss the requirements of astronomers using the network, and some prototype nodes of the VOEventNet that can satisfy these needs.

## 2. Technology

Figure 1 shows the components of the VOEventNet network, which we shall describe below. First we define a Broker to be a service that broadcasts VOEvents from a given domain name and port number. The events may have been received from another broker, or through a publishing system attached to the VTP node (see below). Publication, in the sense of VOEvent, is the assignment of an identifier, and validation of the VOEvent packet. Just as with publishing books, the Publisher interfaces with the Author, which is the entity responsible for the scientific content of the VOEvent; then a closely-coupled broker will actually broadcast the events via VTP.

A broker opens a VTP socket that allows subscribers to connect, and the connection remains open. VOEvents are broadcast to all the subscribers that are currently connected. The Registry (see section 5) will be used to advertise what is available, which brokers are broadcasting which event streams, and documentation of what the data in the events means.

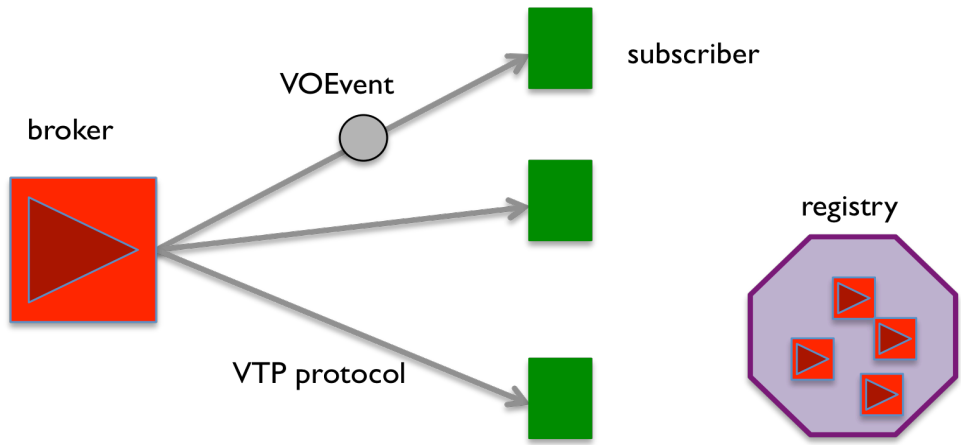

Figure 1: The basic VOEventNet infrastructure. The transport infrastructure is used by brokers to broadcast to subscribers. A global registry will allows discovery of resources such as publishers, brokers, event-stream metadata.

### 2.1 VOEvent

In the past, notices of astronomical transients have been communicated by natural language, but the onset of the deluge means that machines must take much of the burden of decision-making and pointing the telescopes. This realization led to the VOEvent[4][5] standard, which gives authors flexibility in data representation, yet enough structure that subscribers' machines can understand and respond. VOEvent is XML-based, evolved over years, a recommendation of the International Virtual Observatory Alliance[6] (IVOA). The results of astronomical observations using real telescopes are expressed with VOEvent, to be published and transmitted, and then be captured and filtered by subscribers. Each event that survives rigorous filtering can then be passed to other telescopes to acquire real follow-up observations – or passed to computational or archival software for 'virtual follow-ups'. This must happen quickly (often within seconds of the original VOEvent) and must minimize unnecessary expenditures of either real or virtual resources.

A VOEvent packet provides a general purpose mechanism for representing transient astronomical events. The XML schema[7] is as simple as practical to allow the minimal representation of scientifically meaningful, time critical, events. VOEvent also incorporates other standard VO and astronomical schema, specifically STC[8] for space-time coordinates and UCDs[9] to characterize the semantics of the data.

Each event has a unique **identifier** (or IVORN) that is composed of a stream identifier and a 'local' identifier within that stream. The registry is being built so that these identifiers can be resolved: from event identifier to its stream, then finding a repository that has events from that stream, then a query to that repository to resolve the event itself.

In the XML representation of VOEvent, there may be at most one of each of the following optional sub-elements:

- **Who:** This is the contact information for the organization that is responsible for the content of the message.
- **What:** A structured description of the observed parameters, which may include both key-value pairs and tabular data, together with appropriate metadata.
- **WhereWhen:** Space-time coordinates of the event. The data model here is a point in the sky with circular error, and a point in time with an interval of uncertainty. While a wide variety of coordinate systems are possible here, most event authors use ICRS and UTC coordinates.
- **How:** Instrument configuration for the observation.
- **Why:** This is for an initial scientific assessment; a classification and probability.
- **Citations:** This element contains the identifier (IVORN) of other VOEvents that are relevant to the same astrophysical event. A sequence of instrumental observations would each cite one of the others in the chain; follow-up reports cite the IVORN of the discovery event.
- **Description:** This is the place for natural language, a label that can be attached to any part of the rest of the event, describing, for example: one of the parameters, the author organization, etc.
- **Reference:** References should be used for the URLs of related websites.

Only those elements required to convey the event being described need be present. The intent of VOEvent is to represent data associated with an astronomical transient.

The **Who** section of VOEvent allows an event author great flexibility in describing the data that is the heart of the event. While other elements specify facets of the event common to all astronomical transients (spacetime, classifications, provenance, etc.), it is the Who section that allows the event author to create a data model: there are parameters ('Param') that can be organized into Groups, and a Table can be specified with a set of Fields (i.e. table columns). Each such quantity is a number or string, with descriptive metadata, including units, data type, and semantic classification[9]. When the semantic classification is `meta.ref.url`, for example, then the subscriber can assume that it is a URL link.

VOEvents are split into classes called **Streams**, generally based on the instrument or other source of the event. Each event in a stream will have similarly structured Params and Tables (in the What section of the VOEvent), with the same meaning, units, data types, etc. In this way, a subscriber can rely on using one or more specific parameters for building selection criteria. It is considered the responsibility of a Publisher to make sure that each event of a given stream is valid, in terms of the previous metadata (definitions and documentation) that defines the meaning of scientifically relevant data.

VOEvents can be **signed** [10], with a PGP message encoded into the VOEvent. If a Subscriber has the public key of that author, the signature assures the message is from the Author. In terms of security, the VOEventNet has only this assurance of provenance.

There is a **python library** for VOEvent [11] that is built automatically [12] from the VOEvent schema [7], which allows parsing and building of VOEvent XML packets.

## 2.2 VOEvent Transport Protocol

The VOEvent Transport Protocol[11] (VTP) is based on the original GCN protocol[3], in use since 1993, but scaled to wider usage and made for transport of VOEvents. VTP senders include both VOEvent message authors who originate messages and VOEvent brokers which disseminate messages to subscribers. Receivers include both subscribers who use the VOEvent messages and VOEvent brokers which receive messages from Publishers for dissemination to subscribers.

VTP is intentionally as simple as possible while still accomplishing the required task, providing a universal distribution service which supports destination filtering. High-traffic publishers will require some source filtering to prevent flooding of the VOEvent network. All messages are sent over a TCP connection preceded by a 4-byte network-ordered count, followed immediately by the payload data. The 4-byte count is interpreted as a 32-bit integer equal to the number of payload bytes following the count bytes. The payload is considered an opaque collection of bytes at this level, but as described above, all messages are XML documents. No checksum or digest check data is included; the protocol relies on TCP's guaranteed error-free delivery of data. The message receiver makes the TCP connection to the sender of the message. All connections over which a broker sends VOEvent messages are kept open continuously, and 'keep alive' messages are part of the protocol so that broken connections can be detected and remade. The broker periodically sends an 'iamalive' message, to which the subscriber replies with a copy of that message plus some optional identification information.

There are currently three implementations of VTP: Comet[14], Dakota[21], and GCN-TAN[24].

## 2.3 Registry

The VAO[15] is the US member project of the IVOA[6] and aims to provide the necessary infrastructure to support the federation of distributed data sets and services. An event portfolio is the poster child for this type of activity, containing all that can be discovered about an event from different multiwavelength data archives and particular analysis services (annotators). IVOA data access protocols, e.g., ConeSearch, Simple Image Access (SIA) and Table Access (TAP), ensure that the same interface is employed across all data archives, no matter where they are located, to perform the same type of data query. Common data models, e.g., Space-Time Coordinates, Spectral, and TimeSeries, define the shared elements across data and metadata collections and provide a framework for describing relationships between them so that different representations can interoperate in a transparent manner. At the heart of this infrastructure lies the registry, providing a repository for descriptions of all types of astronomical resources, their capabilities and interfaces.

The VOEvent Registry Extension Schema (VOEventRegExt[16]) defines the specific metadata for describing event infrastructure in the registry. The various VOEventNet components outlined above map to one of three resource types: VOEventStream, VOEventServer and VOEventAnnotator. The stream resource is a scientifically coherent collection of events from the same motivation – team, project, or experiment – with each event employing the same vocabulary (parameters) to describe events. The stream resource can also store PGP public keys of valid signatories, so that subscribers can validate the authorship. The server resource describes which computers and interfaces can be used to receive events for dissemination (publish capability), send out future events (subscription capability), and run queries on past events (query capability). It specifies the streams that it knows and the functionality offered for each. Finally, the annotator resource details services that take in events from particular streams, work on certain parameters and produce a specific response in the form of another VOEvent.

A distributed network of registries exists across the IVOA, employing the OAI-PMH protocol[19] to stay in sync with each other. These can be either full registries, holding all resource descriptions, or publishing registries, which manage resource descriptions of a particular provenance, e.g., all those related to a specific project or subject area. The CARNIVORE registry[17] at Caltech will be used as the publishing registry for event infrastructure resource descriptions. Though VOEventNet component metadata will be managed – entered, updated, etc. – through CARNIVORE, it will be discoverable via any registry within the IVOA (see [19] for a list of available registries). In this way, users can easily find where and how to get certain event streams, where to find a persistent copy of a specific event, or details of a particular type or instance of annotation service (see 'Annotate' below).

## 3. An Open VOEventNet

The VTP communication protocol is the basis for the 'VOEventNet', illustrated schematically in Figure 2, which consists of the elements described above, with some extensions to make it more useful. Black lines show the VTP protocol, which allows different event providers to interoperate. Each VTP arrow connects a red broker node to a green subscriber node: the connection is initiated by the subscriber, and the broker broadcasts events to all listeners. The picture shows some other colored symbols and words, these are functions and services that have shown their usefulness in the prototypes – software patterns.

**Publish**: The word here means interfacing with an event author, laying the path for future communication that is automated and fast, yet understandable by others. In practice, every publishing system will have an attached event broker to send out the accepted events in real time to VOEventNet. While an irresponsible publisher may just pass on whatever events any author provides, another might check credentials and only publish events from certain authors; it may check events for VOEvent compliance, and may check the signature for authenticity. More deeply, a publisher may also demand that the corresponding event Stream has been pre-registered, with all metadata provided in advance of the events: units, UCDs, and descriptions. The more responsible a Publisher is, the more Subscribers will trust the content.

**Subscribe**: An entity receiving VOEvents is a Subscriber, meaning they have a connection to a broker, and receiving broadcast. There may also be a human interface that allows creation of custom subscriptions, perhaps the bright transients, or those observable from a given site, or those classified in some way. Customized event selection can be implemented either as custom feeds, or from recognizing the subscriber and using the preferences that have been previously recorded. Note that subscription and query (below) are intimately related in the sense that each is a selection predicate, the former about future events, the latter about past events.

**Relay**: As noted above, a broker can subscribe to other brokers, and then re-broadcast content to subscribers. This pattern could be used for a content aggregator or selector, perhaps taking the 'best' events relevant for a general topic such as supernovae or radio transients, combining multiple publishers. The relay pattern can be used to filter for specific kinds of event, for scheduling on a live robotic telescope. The relay should also be a quality control agent, preventing illegal/malformed VOEvent messages from entering the system or being relayed, and perhaps also informing publishers of any non-compliance.

**Annotate**: This is a combination of subscribing and publishing. Every time an event is received, some new information is added to it, perhaps a follow-up observation, or an archive lookup, or machine intelligence applied to classify the event, and that new information published as a new event that cites the original. It is through annotation that a portfolio is built, containing the multiply-authored VOEvents that together describe an astronomical transient. Some examples of annotators: a service that finds the nearest galaxy to a given transient location; evaluating a light-curve to classify a transient; adding a follow-up observation to the portfolio of a transient.

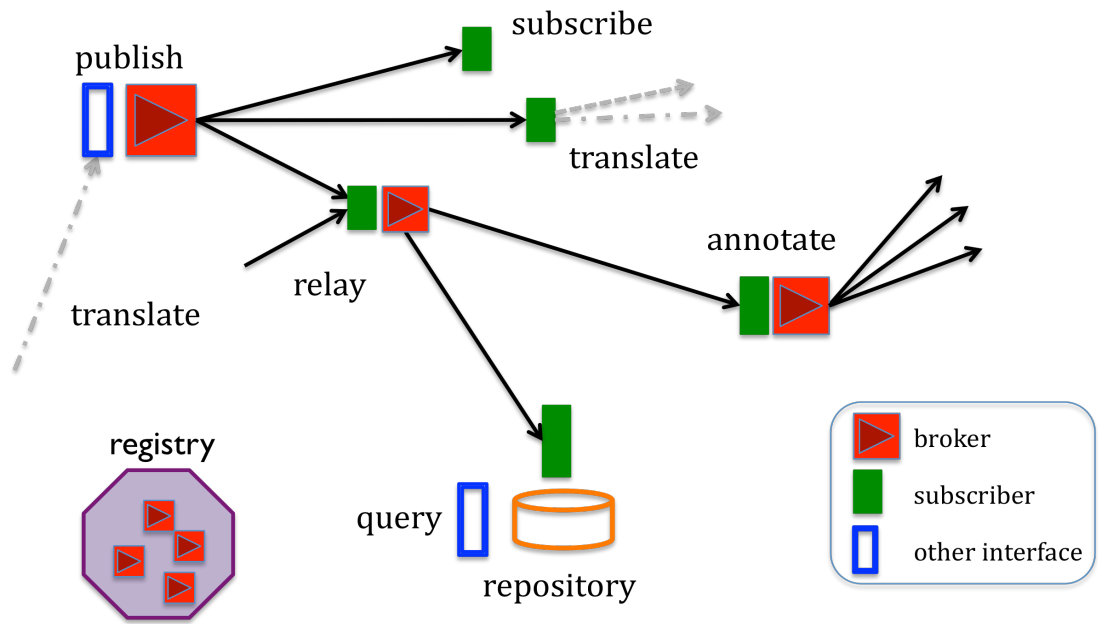

Figure 2: The expanded VOEventNet, with the usage patterns that are explained below. Publish, subscribe, and registry are as before; the blue boxes imply an interaction with users. Translation, relay, and annotate are combinations of these; the repository of events also brings the question of queries and completeness.

**Translate**: There are two kinds of translation implied here: protocol translation and syntax translation. If a fully-formed VOEvent arrives by XMPP, email, http, RSS feed, or inked on the side of a cow, then it is simply a protocol translation; whereas translation to/from other syntaxes is more difficult. Minor-Planet Center notices and natural language GCN Circulars can be converted to VOEvent, and VOEvents can be translated into a human-readable message or a telescope control schedule.

**Repository**: This is a database of past events, which may all come from the same stream, or from different streams, that can be used for coincidence or statistical studies, or to build a training set for supervised classification. The metadata about the meaning of the events (stream metadata) can be cached at the repository, although the universal registry should the authoritative source of this. The repository will ingest some or all of the events that it receives through subscription, and should offer some services and web-forms to allow queries on the event database.

**Query**: We use the word in a restricted sense, as it really means 'selection query'. It is a definition of what is 'interesting' in some sense, it is a selection predicate, a Boolean-valued expression. A query can be used to find past events, and it can be used for future events – in the case that it is used as a selection in subscription. Queries can be expressed in various ways, such as Python, SQL or Xpath expressions, or key-value pairs that are results of web forms.

**Replicate**: We discuss replication of event databases in section 5, which covers future directions.

These are the essential patterns that we expect to become integrated with VOEventNet to form the cyber-infrastructure for astronomical transients.

While we have generally considered public release of event notices in the foregoing, some projects may want a private version of VOEventNet, or a mixed model of public and private event streams. There could be VTP nodes on a private network (intranet), or VOEventNet (public) nodes that use the IP address of the subscriber to decide if a private event can be sent to a given subscriber.

## 4. Existing Software and Services

There are several partners in VOEventNet, with different objectives and different types of events, and yet the shared protocol allows them to interoperate. Following is a short summary of activity and objectives for each.

### 4.1 Dakota

This broker is part of the DC-3 Dreams observatory automation software[20], as well as a dispatch scheduling system for observatories. The dispatch scheduling system includes an automated "closed-loop" VOEvent follow up facility. It receives VOEvent messages, filters them according to user preferences, selects the appropriate user-defined observing strategy, queues and performs follow up observations, and then automatically publishes a follow up VOEvent announcement as well as a web-based portfolio of the data including the imagery and relevant log files. This software has been commercially available since 2009.

Besides its own software products, DC-3 Dreams also developed and open-source published a free cross-platform set of VOEvent tools called Dakota VOEvent Tools[21]. Dakota provides a complete VOEvent backbone connectivity and distribution package for Linux, Mac OS X and Windows systems. Tools consist of a broker/receiver, a command line publisher, and a command line message validator. It serves as a reference implementation of the VOEvent Transport Protocol 1.1, as well as for the VOEvent Digital Signature 1.1 proposal[10]. Dakota is written in the C# language and requires the free Mono libraries on Mac OS X and Linux.

DC-3 Dreams also operates a Dakota VOEvent broker at its facility. This broker is currently configured to connect to NASA GCN-TAN brokers and the VOEvent feed from Caltech Skyalert. Subscribers (other than other brokers) include the AAVSO Bright Star Monitor project and several other amateur and university clients which are running the DC-3 Dreams VOEvent follow up system described above. It is expected that the number of stations will grow quickly once the VOEvent Working Group makes VTP and this architecture its chosen method of connectivity.

### 4.2 GCN/TAN

The original GCN (Gamma-ray Coordinates Network) system[3] has been running since 1993, and publishes all of the astronomical transient notices from the currently operating space-based NASA, ESA, and JAXA observatories, including AGILE, Fermi, Integral, MAXI, and SWIFT, plus the ground-based MOA observatory. It distributes almost 60 different notice types (positions, lightcurves, spectra, and images, current s/c pointing direction, plus test messages) from these missions and instruments in several different formats and transport protocols. (In past there were ~50 notice types from now-defunct missions and instruments.) These notice types cover a large range of transient phenomena. Originally, it was entirely GRB positions in real time (delays ranging from 3-10 sec), but GCN has branched out to include non-GRB transients from unknown sources, flares from blazars, AGN and other known X-/γ-ray sources, flare stars, and gravitational microlensing events. The emphasis is on short-term transient phenomena which need rapid follow-ups by other ground- and space-based instruments in the 10 sec to several hours time range, i.e. automated follow-ups. A complete list of notice types can be found at [22].

GCN/TAN (Transient Astronomy Network) has been recently upgraded to include the VTP publication of VOEvents through three brokers[23]. All three GCN/TAN brokers distribute all the VOEvents that GCN produces (i.e. redundancy). A C-language client is also provided [24]. This client works with all VTP brokers. It is a simple, single file, self-contained client that requires no other software packages to be installed on the client's machine (only a C compiler). Currently, GCN/TAN only distributes VOEvents that have been "ingested" into the GCN system by custom connections from the source institutions (the source-institution custom formats are converted into the standard GCN/TAN VOEvents). The brokers have the ability to publish other event streams (for then further distribution), but this function is disabled because authentication is not yet implemented. Authentication will be implemented within a year.

In addition to the normal anonymous connection by the three VOEvent brokers, which then will receive all the VOEvents that GCN/TAN distributes, subscribers can make prior arrangements with GCN to specify a configuration of which event types they want to receive. This filtering[25] capability is exactly the same as the original GCN filtering capability (by type, location uncertainty, intensity, time delay, identity, and 11 other filtering criteria). The filtering can greatly reduce the number of VOEvents a given site receives. Currently GCN/TAN distributes 2000 VOEvents per day – this number will greatly increase in the next couple of years. Once a filtering configuration is set up, a subscriber connecting to one of the 3 GCN/TAN brokers will automatically be recognized (i.e. "known" site instead of an anonymous site), and then the filtering rules for that site will be used for that site to determine which VOEvents are sent. These filtering configurations can be set up on a GCN/TAN webpage[26].

GCN/TAN has an almost 20-year history of collecting all the astrophysical transient events, converting them to a set of standard output formats, and distributing them by a set of standard protocols to whomever wants to receive them. Notice types have come and gone, and distribution media and protocols have come and gone. It is GCN/TAN policy to aggressively pursue all sources of astrophysical phenomena and publish them into VOEventNet, and we strongly encourage all projects that produce transient phenomena to contact GCN/TAN about this -- science will benefit from the rapid follow-up in other wavebands and particle messengers by other observers.

### 4.3 LOFAR

LOFAR[29] is a new low-frequency radio telescope based in The Netherlands but with stations across Europe. It operates at frequencies between 30 and 240 MHz, combining unprecedented sensitivity and resolution in this regime with an extremely wide field of view. The LOFAR Transients Key Science Project[30] is one of six core scientific projects chosen to inform the design of the telescope and expected to provide the bulk of the initial scientific results.

The design of LOFAR makes it particularly suited for use in both discovering and following up transients. The array consists of a large number of static antennae, each of which is sensitive to the whole sky above it. Signals are digitized and combined in software to "point" the telescope in a particular direction. Indeed, the way in which signals can be combined is limited only by the available computing resources and network bandwidth: given sufficient resources, multiple areas of sky (or "beams"), may be observed simultaneously. Depending on frequency, each beam can have a field of view of hundreds of square degrees. LOFAR will regularly tile out a large fraction of the sky with multiple beams, identifying and recording all transient and variable sources at cadences as fast as 1 second. This will result in a light-curve archive that is predicted to grow at a rate of up to 50 TB/year. Furthermore, notifications of noteworthy events will be distributed to the community.

Since LOFAR is completely software driven, repointing the telescope can happen very quickly: there is no need to wait for mechanical moving parts to spring into action. This means that LOFAR can respond exceptionally quickly to targets of opportunity. Going even further: in some observing modes, the raw data received by each dipole is buffered for a short period (on the order of tens of seconds; it is possible to trade off bandwidth for time). On the receipt of an appropriate trigger, the buffered data can be read out to disk. It is then possible to make an image in any direction – even if the telescope was pointing somewhere else – at any time included in the buffer. In other words, if notification of a new transient can be delivered quickly enough, LOFAR can act as a "time machine", observing the location of the transient **before it happens**!

The LOFAR Transients Key Science Project has committed to deliver notifications of new LOFAR transients to the community using VOEventNet, and will similarly monitor VOEventNet for notifications of transients detected by other facilities and follow-up where appropriate. The relevant infrastructure at LOFAR is, at time of writing, actively under construction; this includes the Comet VOEvent Broker[12]. We anticipate the first LOFAR transients being announced by VOEvent before the end of 2012.

Two projects closely related to LOFAR also bear mentioning. **4 PI SKY**[31] builds upon the VOEvent infrastructure being developed for LOFAR to coordinate response to astronomical transients across the SKA pathfinder telescopes (LOFAR, ASKAP and MeerKAT), as well as their partners in other wavelength regimes. Between them, these telescopes provide true all sky (4 $\pi$ steradian) coverage.

Meanwhile, **AARTFAAC**[32] is extending the core part of LOFAR to provide a 24/7 all visible sky transient monitor. This will provide a unique opportunity to observe the rarest events, notification of which will be rapidly disseminated by VOEvent.

**4.4 Skyalert**

Skyalert[31] is a clearinghouse and repository of information about astronomical transients, each described by a collection of VOEvent packets, with both web page and web service interfaces. It is a way to browse recent and past transients, as tables, multi-layered web pages, or with popular astronomical software such as Worldwide Telescope[32]. Events are displayed on the front page[31] on a time line that emphasizes freshness: the events from the recent hours are prominent. Skyalert receives events through several protocols, and in some cases translates from another (non-VOEvent) representation. Some available event streams are:

- AAVSO special notices
- CBAT supernovae
- MOA and OGLE microlensing surveys
- GCN-brokered feeds, including AGILE, Fermi, INTEGRAL, MAXI, SWIFT
- GCN circulars as annotation, translated from email
- CRTS worldwide survey, with classification annotation and follow-ups from Palomar
- CSS NEO discoveries, translated from email
- Pi of the Sky transients

There is a web-based **subscription** system, so that information about transients can be delivered to users and their telescopes immediately upon receipt. Subscriptions can pick carefully by inviting an arbitrarily complex python expression on the values of parameters, to decide if a portfolio is interesting. Thus system-created and user-created feeds are created; as VTP broadcast, email, or other messages.

Authors can work with Skyalert as a **publisher**: the formal definition of a stream is through a *sample event*, where all the parameters that might be used (in a real event) are present in the sample, with full metadata. Once it is built, Skyalert staff bring the new stream online, where the author can evolve it by web forms, adding descriptive content and new parameters. The author's username and password is then used in a web form or web service to publish each new event.

Skyalert provides an event **repository**, storing all events that come through the broker, and allowing bulk queries and drill-down. The query and subscription system are unified: selection of future events through subscription uses the same Python Boolean expression as selection of past events as a query. The repository can also be queried through a simple web-based API; examples are in the Skyalert Client Kit[33]. This paradigm of 'click or code' runs through the Skyalert development, that whatever is available through the web pages is also available through a web service API.

Skyalert adds the concept of **portfolio** to the VOEvent vocabulary: it is the collection of VOEvents that supports the hypothesis that that come from the same object. A portfolio might consist of an initial observation, a follow-up observation, an event containing a classification, and an event containing archival information about that spot in the sky. Each event can be differently authored, so long as one cites the other. The Skyalert visualization of a transient includes all the VOEvents in its portfolio.

There is also a **security** infrastructure provided by Skyalert: each user logs in with username and password, and each event stream has one of these as its owner, so that publishing events can only be done by this user. For viewing events, some streams are accessible only to a given group of users – similar to a Unix group – and the Skyalert staff then assign users to groups to give that access.

Skyalert is released as open-source **software**[34] to allow local implementations as well as the web-based application. It requires Python and the web framework Django. Hence a development platform for building real-time decision rules about transients, and for mining the repository. As noted above, the Skyalert project also maintains the VOEventLib[11] software for validating, reading, modifying, and writing VOEvent packets.

As with GCN/TAN, it is Skyalert policy to aggressively pursue all sources of astrophysical phenomena and publish them into VOEventNet, and we strongly encourage all projects that produce transient phenomena to contact Skyalert about this.

## 5. Future Directions

We hope and expect that VOEvent and VOEventNet will become part of the information infrastructure for rapid distribution and understanding of astronomical transients. Some directions for this progress are summarized below.

**Registry implementation**

The Virtual Astronomical Observatory project is committed to full implementation of the registry structure, for discovery and utilization of VOEvent streams and servers, as described above.

**Scalability and software ecosystem**

As event producers come online, the numbers of events will grow from a few per day to thousands per day. Brokers, subscribers, and repositories will need to not just handle, but judge and select, the important events. To make this effective, we need multiple, robust implementations of the VTP, as well as specialized nodes of VOEventNet for specialized functions, such as telescope control, repository systems, machine intelligence nodes, and rapid decision nodes.

**Database replication**

A centralized event database is not scalable when the number of events, streams, and subscribers becomes a deluge. Rather we expect most large projects will build their own repository, and small projects will use a general facility such as Skyalert. However, we would like to emphasize the idea of replication of important streams at multiple sites, keeping the collections synchronized with low latency. Given that many existing VOEvent streams have URL-valued parameters, we can think of replicating the data objects that the URLs point to; in this case, note that the VOEvents in the replicated repository will not be identical to the originals, as these URLs will have been substituted. Shown in Figure 2 is a dashed line that may in the future become a standard: perhaps based on the OAI-PMH protocol[19] that the digital library community – and the Virtual Observatory – uses for replicating metadata.

**Broadening the community**

We expect and hope that VOEventNet will become the dominant cyberinfrastructure for machine-understandable astronomical transients: for that to happen, it needs to be straightforward for an event author to publish into VOEventNet, and easy for potential subscribers to discover that stream. Amateur astronomers will be encouraged to make scientifically-relevant observations, for example discovering Near Earth Asteroids, following up microlensing events, or following up GRBs through slaving their telescope to an orbiting observatory such as SWIFT. Large telescopes can also be activated through VOEventNet for follow-up of important events from facilities such as LIGO or LSST.